\documentclass[article,12pt]{elsarticle}

\usepackage[T1]{fontenc}
\usepackage[utf8]{inputenc}
\usepackage{lmodern}
\usepackage{microtype}
\usepackage{amsmath,amssymb}
\usepackage{graphicx}
\usepackage{xcolor}
\usepackage{booktabs}
\usepackage{tabularx}
\usepackage[hidelinks]{hyperref}

\newcommand{\Qbeta}{\texorpdfstring{Q$\beta$}{Q-beta}}
\hypersetup{
  pdftitle={Bacteriophage \Qbeta{}: Six Decades at the Frontier of Molecular and Viral Evolution},
  pdfauthor={Ester Lázaro and Susanna Manrubia}
}

\biboptions{sort&compress}

\makeatletter
\renewcommand{\emailauthor}[2]{%
  \stepcounter{ead}%
  \g@addto@macro\@elseads{\raggedright%
    \def\@@tmp{#1}%
    \eadsep{\ttfamily\expandafter\strip@prefix\meaning\@@tmp}%
    \def\eadsep{\unskip,\space}}%
}
\makeatother

\journal{Current Opinion in Virology}

\begin{document}

\begin{frontmatter}

\title{Bacteriophage \Qbeta{}: Six Decades at the Frontier of Molecular and Viral Evolution}

\author[aff1]{Ester Lázaro\corref{cor1}}
\ead{lazarole@cab.inta-csic.es, smanrubia@mncn.csic.es}
\author[aff2,aff3]{Susanna Manrubia\corref{cor1}}

\cortext[cor1]{Corresponding authors.}
\address[aff1]{Centro de Astrobiología (CAB), CSIC-INTA, Ctra. de Ajalvir km. 4, 28850 Torrejón de Ardoz, Madrid, Spain}
\address[aff2]{Museo Nacional de Ciencias Naturales (CSIC), c/ José Gutiérrez Abascal 2, 28006 Madrid, Spain}
\address[aff3]{Grupo Interdisciplinar de Sistemas Complejos (GISC), Madrid, Spain}


\begin{abstract}
Bacteriophage Q$\beta$, first isolated in 1961, has served for more than six decades as a model system for molecular and viral evolution. The development of cell-free replication systems based on the \Qbeta{} replicase enabled the first demonstrations of Darwinian evolution in RNA molecules. In parallel, the phage's compact RNA genome, high mutation rate, and ease of propagation established \Qbeta{} as an exceptionally tractable model for studying viral evolution. The discovery of extensive genetic heterogeneity made of \Qbeta{} the first biological system in which molecular quasispecies theory found direct experimental support. Building on this foundation, decades of evolution experiments under thermal, ecological, and mutagenic pressures established how mutation, selection, epistasis, and population structure jointly shape adaptive trajectories. Deep sequencing has recently allowed the mutant spectrum of Q$\beta$ populations to be resolved as large, structured genotype networks, revealing a hierarchical organization around dominant sequences that generalizes classical mutation-selection balance to a dynamic process, and uncovering topological features shared with RNA viruses evolving within natural hosts. In parallel, the catalytic core of the Q$\beta$ replication system has been repurposed in synthetic biology to revive cell-free Darwinian evolution within compartmentalized, cell-like systems, opening new avenues for studying the origin and early evolution of life. Further recent work has pushed the phage into new experimental territories, including extreme, astrobiologically relevant environments and biotechnological platforms for peptide display. Together, these developments illustrate how a single bacteriophage has repeatedly anticipated and continues to clarify general principles of molecular, population, and ecological evolution.
\end{abstract} 


\end{frontmatter}


\section{Introduction}

Bacteriophages have long served as foundational model systems in molecular biology. Research with phages has contributed to major advances, such as the demonstration that DNA carries genetic information, the resolution of its double-helix structure or the identification of messenger RNA.
In addition, studies of phage-host interactions have led to the development of techniques that underpin modern biology, such as CRISPR-Cas phage resistance systems \cite{salmond:2015}. Among them, bacteriophage \Qbeta{} occupies a unique place at the intersection of molecular biology and evolutionary virology.

First isolated in 1961 by Ichiro Watanabe and colleagues from sewage samples in Japan~\cite{watanabe:1967}, \Qbeta{} is a non-enveloped positive-sense single-stranded RNA (+ssRNA) virus with an icosahedral capsid of approximately 28~nm. Its genome comprises about 4217 nucleotides and encodes four proteins: the coat protein, the readthrough protein A1~\cite{hofstetter:1974}, the maturation/lysis protein A2~\cite{bernhardt:2001,cui:2017}, and the $\beta$ subunit of the viral RNA-dependent RNA polymerase (replicase) ~\cite{blumenthal:1979}. The functional enzyme is a heterotetrameric complex formed through the assembly of the phage $\beta$ subunit with three host \emph{Escherichia coli} proteins: ribosomal protein S1 and elongation factors EF-Tu and EF-Ts. Notably, the \Qbeta{} replicase lacks proofreading activity and exhibits one of the highest mutation rates measured for RNA viruses, estimated at approximately $1.4\times10^{-4}$ substitutions per nucleotide per round of copying~\cite{bradwell:2013}. Under the current taxonomy of the International Committee on Taxonomy of Viruses, \Qbeta{} belongs to the family \emph{Fiersviridae} and is classified as the species \emph{Qubevirus durum} within the genus \emph{Qubevirus}~\cite{walker:2021}.

Several biological features of the virus, including its compact genome, high mutation rate, short generation time, and ease of propagation under laboratory conditions, facilitated experimental investigations of viral adaptation. Over subsequent decades, \Qbeta{} evolved from a model for studying RNA replication and quasispecies dynamics into one of the most versatile systems for experimental evolution. Although it infects bacteria, many of the principles uncovered through studies performed with \Qbeta{} are directly relevant to the evolution of RNA viruses in general, including medically important pathogens. In this review, we summarize how studies using \Qbeta{} have shaped our current understanding of viral quasispecies, molecular evolution, and adaptive processes operating across molecular, population, and ecological scales. Major breakthroughs are summarized in Table \ref{tab:qbeta-timeline}.

\begin{table}[!htbp]
\caption{Main milestones in the study of bacteriophage \Qbeta{} as a model
  system. Blocks A to E are ordered by the date of their first milestone and, within
  each block, chronologically.}
\label{tab:qbeta-timeline}
\scriptsize
\renewcommand{\arraystretch}{1.1}
\setlength{\tabcolsep}{4pt}
\begin{tabularx}{\textwidth}{@{}%
  >{\raggedright\arraybackslash}p{0.13\textwidth}%
  >{\raggedright\arraybackslash}X%
  >{\raggedright\arraybackslash}p{0.41\textwidth}@{}}
\toprule
Year(s) & Milestone & Scientists and references \\
\midrule

\multicolumn{3}{@{}l}{\textcolor{blue}{\textbf{A. Isolation, replicase and cell-free evolution}}}\\
1961 & Isolation of \Qbeta{} from sewage &
  Watanabe et~al.~\cite{watanabe:1967}; Sakurai et~al.~\cite{sakurai:1968} \\
1964--1965 & Template-specific replicase; RNA replication \emph{in vitro} &
  Haruna \& Spiegelman~\cite{haruna:1964,haruna:1965} \\
1965 & Self-propagating, infectious RNA made \emph{in vitro} &
  Spiegelman et~al.~\cite{spiegelman:1965} \\
1967 & Cell-free Darwinian evolution (``Spiegelman's Monster'') &
  Mills, Peterson \& Spiegelman~\cite{mills:1967} \\

\addlinespace[0.2em]
\multicolumn{3}{@{}l}{\textcolor{blue}{\textbf{B. Molecular genetics and the quasispecies concept}}}\\
1971--1977 & Quasispecies theory, error threshold &
  Eigen~\cite{eigen:1971}; Eigen \& Schuster~\cite{eigen-schuster:1977} \\
1974--1976 & Site-directed mutagenesis of \Qbeta{} RNA &
  Flavell et~al.~\cite{flavell:1974}; Domingo et~al.~\cite{domingo:1976} \\
1978 & First experimental evidence of a quasispecies &
  Domingo et~al.~\cite{domingo:1978} \\
1981 & \emph{De novo} synthesis of self-replicating RNA &
  Biebricher et~al.~\cite{biebricher:1981} \\

\addlinespace[0.2em]
\multicolumn{3}{@{}l}{\textcolor{blue}{\textbf{C. Viral adaptation and evolutionary dynamics}}}\\
2008--2016 & Adaptation under increased mutational load &
  Cases-González et~al.~\cite{cases-gonzalez:2008};
  Cabanillas et~al.~\cite{cabanillas:2013,cabanillas:2014};
  Arribas et~al.~\cite{arribas:2016} \\
2010--2018 & Virion thermal stability and trade-offs &
  Domingo-Calap et~al.~\cite{domingo-calap:2010};
  Garc\'ia-Villada \& Drake~\cite{garcia-villada:2013};
  Lázaro et~al.~\cite{lazaro:2018} \\
2011 & Host--virus coevolution &
  Kashiwagi \& Yomo~\cite{kashiwagi:2011} \\
2013 & Mutation rate measured ($\sim$$10^{-4}$/nt) &
  Bradwell et~al.~\cite{bradwell:2013} \\
2014--2022 & Genetic basis of thermal adaptation &
  Arribas et~al.~\cite{arribas:2014,arribas:2018};
  Arribas \& L\'azaro~\cite{arribas:2021};
  Kashiwagi et~al.~\cite{kashiwagi:2014,kashiwagi:2018};
  Somovilla et~al.~\cite{somovilla:2019,somovilla:2022};
  Hossain et~al.~\cite{hossain:2020} \\
2022--2025 & Host density shapes adaptive dynamics &
  Laguna-Castro \& L\'azaro~\cite{laguna-castro:2022};
  Laguna-Castro et~al.~\cite{laguna-castro:2023,laguna-castro:2025} \\

\addlinespace[0.2em]
\multicolumn{3}{@{}l}{\textcolor{blue}{\textbf{D. Synthetic biology and compartmentalized RNA replication}}}\\
2008 & Self-encoding RNA replication in liposomes &
  Kita et~al.~\cite{kita:2008} \\
2013 & $>$600 generations of compartmentalized evolution &
  Ichihashi et~al.~\cite{ichihashi:2013} \\
2018 & Coevolving, cooperating RNA replicators &
  Mizuuchi \& Ichihashi~\cite{mizuuchi:2018} \\
2020 & Host--parasite RNA ecosystems &
  Furubayashi et~al.~\cite{furubayashi:2020} \\

\addlinespace[0.2em]
\multicolumn{3}{@{}l}{\textcolor{blue}{\textbf{E. Engineering, extreme environments and sequence-space networks}}}\\
2011--2021 & A1-based surface display of foreign peptides &
  Singleton et~al.~\cite{singleton:2018};
  Nchinda et~al.~\cite{nchinda:2021} \\
2024--2026 & Survival under astrobiologically relevant stress &
  Laguna-Castro et~al.~\cite{laguna-castro:2024};
  Rodríguez-Moreno et~al.~\cite{rodriguez-moreno:2025};
  Arribas Tiemblo et~al.~\cite{arribas-tiemblo:2026} \\
2026 & Genotype networks from deep sequencing; generic network features &
  Seoane et~al.~\cite{seoane:2026} ;
  Martínez-Alcalá et~al.~\cite{martinezalcala:2026};
  Manrubia et~al.~\cite{manrubia:2026} \\

\bottomrule
\end{tabularx}
\end{table}

\section{Early experiments with \Qbeta{}}

\subsection{Molecular evolution}

Shortly after its isolation, the experimental potential of \Qbeta{} was greatly expanded by the development of cell-free replication systems based on its replicase~\cite{haruna:1965}. Reconstitution of the enzyme with its three host cofactors enabled experimental studies of RNA replication and, ultimately, laid the groundwork for experiments designed to demonstrate Darwinian evolution in cell-free RNA populations.

In the early-1960s, Spiegelman was interested in understanding self-replication and evolution at the molecular level. The \Qbeta{} phage physically reached his lab in 1963 via Ichiro Haruna, a close associate of Watanabe's academic network, who brought the viral strains with him upon joining Spiegelman's lab as a postdoctoral researcher. After demonstrating that the purified RNA-dependent RNA polymerase induced by the \Qbeta{} phage was strictly template-specific~\cite{haruna:1964}, Spiegelman's group combined \Qbeta{} genome in a test tube with its own replicase, nucleotides, and salts, and propagated it by repeatedly transferring small aliquots into fresh reaction mixture~\cite{spiegelman:1965,mills:1967}. After 74 transfers, a short sequence, 218 nucleotides long, with the ability to replicate with high efficiency had been generated. The result of the experiment was shocking at the time, so much so that the final sequence became known as ``Spiegelman's Monster.'' Building on Spiegelman's approach, later experiments using a combination of HIV-1 reverse transcriptase and T7 RNA polymerase yielded a similar evolutionary system in which RNA molecules corresponding to a 220-base segment of the HIV-1 genome ultimately reduced their size to just 48 or 54 nucleotides, approaching the minimal functional size compatible with sustained replication~\cite{oehlenschlager:1997}.

In the \emph{in vitro} environment, many of the selective pressures experienced by the phage differed from those in the cellular environment. Though the implications of a high mutation rate in the phage were not clearly understood at the time, error-prone replication continuously generated populations of competing variants. Selection then acted primarily on replication speed, giving shorter RNAs an advantage. During serial transfer, shorter and more efficient variants regularly emerged and outcompeted longer ones, resulting in a continuous turnover in which intermediate variants could not fix before new ones arose. Importantly, {\it in vitro} evolution was not limited to the selection of progressively shorter and faster-replicating molecules. The introduction of additional selective pressures, including ethidium bromide and RNA-degrading activities, led to the emergence of variants with improved replication under adverse conditions \cite{saffhill:1970}, demonstrating that cell-free RNA populations can adapt to environmental challenges in ways analogous to biological organisms. Altogether, the experiments showed that evolution could be observed directly in the laboratory on a timescale of hours to days.

In G\"ottingen, Manfred Eigen was thinking about molecular self-organization and the origin of life: how could early self-replicators, lacking proofreading, retain genetic information without either degrading under excess mutation or stalling for lack of it? \footnote{Driven by conversations surrounding Sol Spiegelman's landmark ``test-tube evolution'' experiments, Eigen established a direct scientific exchange with Spiegelman's laboratory and acquired the standard Q$\beta$ bacteriophage stocks and its isolated, template-specific RNA-dependent RNA polymerase. Among others, this system allowed Eigen and collaborators to systematically unravel the kinetics of RNA replication and demonstrate the {\it de novo} synthesis of self-replicating nucleic acids~\cite{biebricher:1981}.} Prompted by a lunchtime suggestion from Francis Crick to formalize the problem mathematically, Eigen built a population-dynamics model of replicating and mutating molecules; in 1971, he defined the quasispecies as its stationary solution~\cite{eigen:1971}. In the framework of Eigen's theory, a quasispecies is no longer a single genotype (or a collection of independent genomes) but a dynamic distribution of related variants maintained by mutation and replication, typically centered around a master sequence. The model also predicted a genotypic error threshold beyond which no stable genetic information can persist.\footnote{Eigen's formulation assumed the absence of reversions, a fitness-peak landscape, and a non-redundant genotype-to-phenotype map. When these conditions are relaxed to mimic more realistic quasispecies, the loss of the master sequence does not imply the loss of function or biological viability of the quasispecies. The genotypic error threshold is a formal result with weak implications for natural populations~\cite{schuster:1999,manrubia:2010} that, nonetheless, has been inspiring in devising strategies to cause viral extinction~\cite{swanstrom:2022}.} The quasispecies grows as a collective unit rather than as a set of independent sequences, and is itself the object of selection~\cite{eigen:1971,eigen-schuster:1977,biebricher:2006}. 

Compact and experimentally tractable, \Qbeta{} provided a system in which the high genetic heterogeneity expected in early self-replicators would be first empirically observed.

\subsection{\Qbeta{} as a molecular quasispecies}

The development of site-directed mutagenesis in Charles Weissmann's laboratory provided a complementary way to investigate the genetic basis of Q$\beta$ replication. During his postdoctoral fellowship in the laboratory of Nobel Laureate Severo Ochoa at New York University---a research group deeply interested in the biosynthesis and replication of nucleic acids---Weissmann \cite{borst:2026} acquired the Q$\beta$ bacteriophage from Spiegelman's laboratory to resolve ongoing controversies surrounding viral RNA replication \cite{weissmann:2011}. Upon his appointment to the University of Z\"urich in 1967, Weissmann took his established Q$\beta$ viral lines and purified enzymes with him, laying the groundwork for his institute's pioneering experiments on site-directed mutagenesis.

Weissmann and colleagues devised methods to introduce mutations at predetermined positions in the Q$\beta$ genome, initially to examine the functional constraints on sequences required for recognition by the viral replicase~\cite{flavell:1974,domingo:1976}. Among them, Esteban Domingo, then a postdoctoral researcher in Weissmann's laboratory, undertook the analysis of individual mutant clones---an effort that led to the unexpected finding that apparently homogeneous Q$\beta$ populations were far from genetically uniform. This heterogeneity pointed to a population maintained in a dynamic equilibrium, with new mutants arising at a high rate and being continually counterselected~\cite{domingo:1978}. Thus, rather than possessing a single defined genome, the phage population could 
be described as a weighted ensemble of closely related sequences, thus providing the first experimental evidence of the genetic heterogeneity underlying the quasispecies picture.

Beyond this molecular setting, RNA viruses have been repeatedly claimed to be actual examples of complex mutant distributions~\cite{domingo:2012,andino:2015,domingo:2016}. However, the extension of the concept of quasispecies to viruses introduced additional layers of complexity. Though the term ``viral quasispecies'' is still broadly applied to viral populations, the complexity of the latter separates theoretical results from Eigen's molecular quasispecies theory from viral dynamics and organization. First, viral populations are finite and frequently experience population bottlenecks, genetic drift, and fluctuating selective pressures~\cite{domingo:2012,wilke:2005}, whereas the original theoretical treatment assumed effectively infinite populations evolving under constant conditions. Second, viral populations rarely attain the mutation--selection equilibrium in which molecular quasispecies are formally defined, because ongoing environmental changes, host adaptation, and demographic fluctuations continuously modify the selective landscape. Viral variants replicate within cells and may interact through competition, complementation, and cooperation, allowing population-level properties to emerge from interactions among genomes~\cite{vignuzzi-lopez:2019,vignuzzi:2006,grande-perez:2005,leeks:2023}. Consequently, viral evolution cannot be fully understood by analyzing individual genotypes in isolation but requires consideration of the mutant spectrum as an integrated and dynamic element.

\section{\Qbeta{} as a model system to dissect viral adaptation}

Beyond its historical contribution to the development of quasispecies theory, \Qbeta{} has also emerged as a powerful experimental system to investigate the mechanisms of viral adaptation. Evolution experiments conducted under a wide variety of selective pressures have revealed how mutation, selection, epistasis, ecological interactions, and environmental heterogeneity jointly shape evolutionary trajectories. 

\subsection{Thermal evolution as a framework to study adaptation}

Many experimental evolution studies using \Qbeta{} have employed temperature as a broad-spectrum selective pressure. Beyond demonstrating the remarkable adaptive potential of the virus~\cite{kashiwagi:2018}, these experiments provided a framework for addressing fundamental questions in evolutionary biology, including the contribution of standing genetic variation and {\it de novo} mutations to adaptation~\cite{somovilla:2022}, the importance of synonymous mutations~\cite{kashiwagi:2014}, and the existence of multiple genetic routes to similar fitness outcomes~\cite{hossain:2020}. They also revealed the influence of clonal interference, epistasis, population structure, and mutation-selection dynamics on adaptive trajectories~\cite{arribas:2021,seoane:2026}, showing that evolutionary success depends not only on the fitness effects of mutations but also on the population context in which they arise~\cite{arribas:2018}.

Studies using abrupt, gradual, and fluctuating thermal challenges further demonstrated how environmental complexity shapes evolutionary outcomes ~\cite{somovilla:2019,arribas:2014}. Experiments targeting the extracellular phase of the viral life cycle showed that \Qbeta{} can simultaneously improve replication at elevated temperatures and virion thermostability, with only limited evidence for strong trade-offs between survival and reproduction ~\cite{lazaro:2018,domingo-calap:2010}, although other selection experiments revealed trade-offs between fecundity and lifespan ~\cite{garcia-villada:2013}. 

Temperature is, however, only one example of a selective pressure that can be applied in a controlled, unidirectional manner; other experiments have instead exposed \Qbeta{} to selective landscapes that change dynamically as a result of the virus's own interactions with its host.

\subsection{Adaptation in dynamic and mutagenic environments}

Coevolution experiments with \emph{Escherichia coli} demonstrated that reciprocal adaptation between host and virus can generate rapid and sustained evolutionary change, producing dynamic fitness landscapes in which the targets of selection are themselves evolving~\cite{kashiwagi:2011}. More recent work expanded this ecological perspective by showing that host population density strongly modulates viral adaptive dynamics, altering infection opportunities, the intensity of selection, and the relative advantage conferred by different mutations~\cite{laguna-castro:2022,laguna-castro:2023,laguna-castro:2025}.

Other experiments using mutagenic nucleoside analogues revealed that adaptation to increased mutational loads may proceed through alternative mutations in the viral replicase~\cite{cabanillas:2013,cabanillas:2014} and that elevated mutation rates can sometimes facilitate adaptation by increasing genetic diversity~\cite{cases-gonzalez:2008,arribas:2016}. Together, these works established \Qbeta{} as a powerful model for investigating both the drivers and limits of viral adaptation~\cite{bull:2004}, highlighting the interplay between ecology, mutation, robustness, and evolvability in determining viral evolutionary outcomes.

\section{Quasispecies characterization through deep sequencing}

\begin{figure}[h!]
  \centering
  \includegraphics[width=\textwidth]{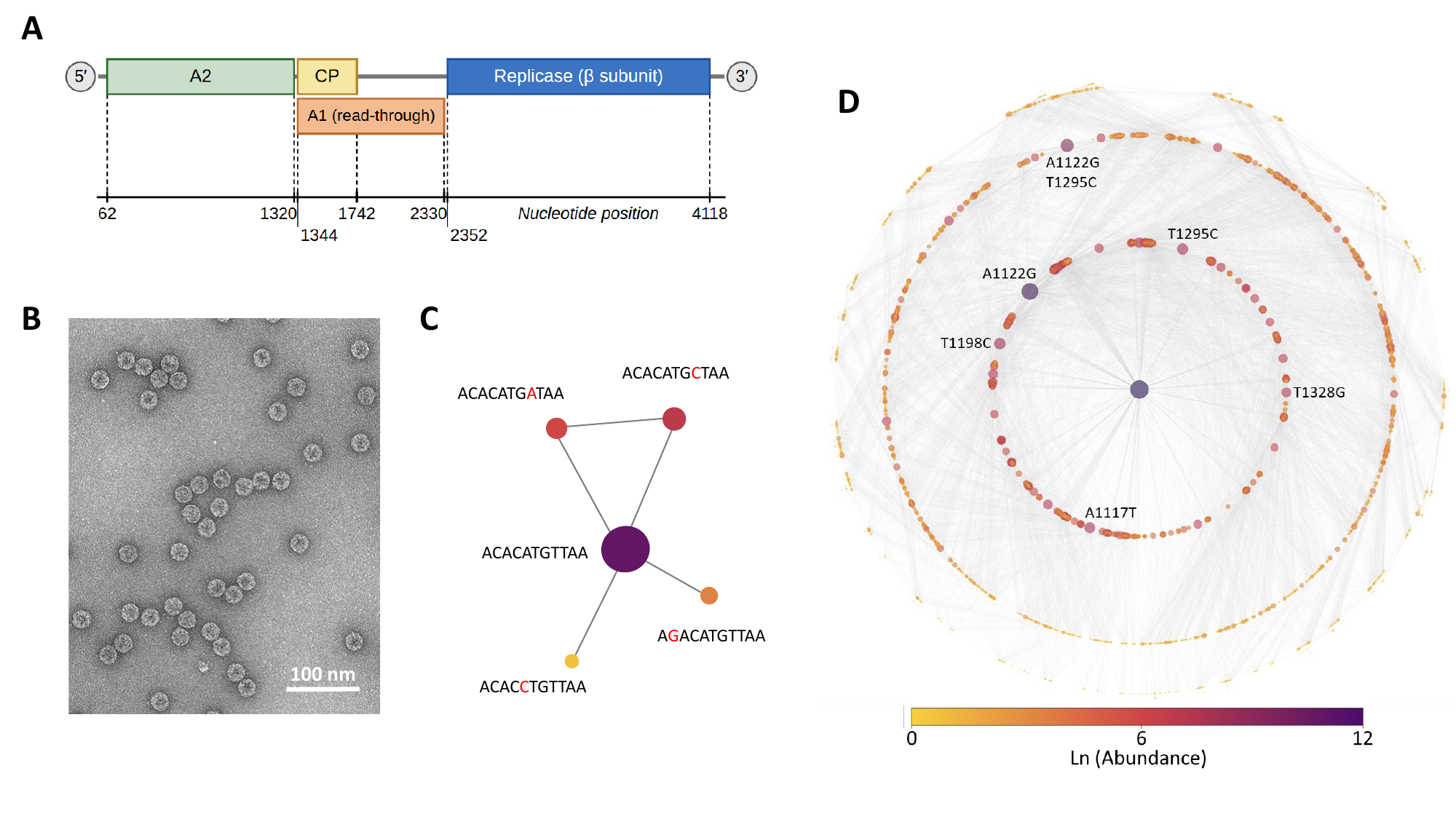}
  \caption{{\bf Genome, particle and mutational landscape of bacteriophage Q$\beta$.}
\textbf{(A)}~Organization of the Q$\beta$ genome, drawn to scale; numbers indicate nucleotide positions.
The genome encodes the maturation protein A2, the coat protein (CP), the minor coat protein A1, which results from read-through of the CP stop codon, and the $\beta$~subunit of the replicase.
\textbf{(B)}~Negative-stain transmission electron micrograph of Q$\beta$ particles. Image courtesy of the Electron Microscopy Service of the Centro Nacional de Biotecnolog\'ia (CNB-CSIC); photo by Mara Laguna.
\textbf{(C)}~Construction of a genotype network. Sequences (here, an illustrative 11-nucleotide window) are represented as nodes, and two nodes are joined by an edge when they differ at a single nucleotide (red). Node size and color indicate abundance.
\textbf{(D)}~Genotype network reconstructed for a Q$\beta$ population adapted for 60 passages to 37$^\circ$C. Data correspond to deep-sequencing of an amplicon spanning from position 1060 to 1330. The central node is the most abundant haplotype, and concentric rings group variants by their mutational distance from it. Node size and color represent the natural logarithm of variant abundance. Selected variants are labeled by their nucleotide substitutions. Additional information can be found in the Supplementary Material of \cite{seoane:2026}.}
  \label{fig:network}
\end{figure}

Deep sequencing has transformed the study of viral quasispecies by making it possible to sample the mutant spectrum of a population at read depths orders of magnitude beyond what individual clone sequencing could achieve \cite{beerenwinkel:2012,isakov:2015}. This technique has allowed, among others, the reconstruction of a near-complete single-nucleotide fitness landscape for an RNA virus \cite{acevedo:2014} and the identification of mutant cloud complexity as an epidemiologically evolvable trait driven by host-pathogen factors \cite{martinezgonzalez:2025}.

This depth of sampling allows a much more extensive census of the mutant spectrum, at read depths unthinkable in the 1970s, and made it possible to ask not just how diverse a \Qbeta{} population is, but how that diversity is structured in sequence space: this latter goal has been recently addressed through the reconstruction of genotype networks (Figure~\ref{fig:network}). Data from an amplicon spanning the \Qbeta{} A2 protein was used to reconstruct genotype networks comprising tens of thousands of haplotypes connected by single mutations \cite{seoane:2026}. Consistently across populations and environments, this reconstruction revealed a hierarchical organization in which the population exhaustively sampled the mutational neighborhood of its dominant sequence while maintaining a periphery of rare variants. This picture---local, exhaustive exploration around a shifting master sequence rather than a static equilibrium---generalizes the mutation--selection balance envisioned by Eigen to a dynamic, structured process. 

This network-based approach is currently being extended beyond the controlled laboratory setting to natural populations. Comparisons of \Qbeta{} genotype networks with those of SARS-CoV-2 evolving within human hosts revealed shared topological features despite the very different biology and selective environments of the two viruses \cite{martinezalcala:2026}. This convergence suggests that the network architecture of a viral quasispecies, first resolved in \Qbeta{}, might be a generic feature of RNA virus populations rather than a peculiarity of one experimental system---consistent with a broader picture in which the diffusive exploration of sequence space by replicating populations follows shared physical principles across very different genomes and hosts \cite{manrubia:2026}.

\section{\Qbeta{} beyond viral adaptation: new experimental platforms}

\subsection{Synthetic biology: reviving cell-free Darwinian evolution}

Beyond its role in dissecting mutation and adaptation in the intact virus, \Qbeta{} replicase has more recently been repurposed as the catalytic core of synthetic self-replicating systems that revisit, in a compartmentalized setting, the cell-free Darwinian evolution first demonstrated by Spiegelman. Kita et al. reconstituted a ``self-encoding system'' inside liposomes, in which the catalytic $\beta$ subunit of \Qbeta{} replicase, translated in situ from an encapsulated template RNA using a purified translation system, in turn replicates that same template ---restoring, within a lipid compartment, the coupling between genotype and phenotype that Spiegelman's naked in vitro system lacked \cite{kita:2008}. Building on this platform, sustained Darwinian evolution of the encapsulated RNA over 600 generations of compartmentalized replication and translation was demonstrated, with genomic RNA progressively reinforcing its interaction with the translated replicase \cite{ichihashi:2013}. Subsequent work extended the system to the evolution of cooperation between coexisting RNA replicators \cite{mizuuchi:2018} and to the emergence and diversification of host-parasite RNA ecosystems \cite{furubayashi:2020}. These synthetic constructs show how the enzymatic machinery that first enabled Spiegelman's experiments continues to serve as a tractable model for questions in the origin and early evolution of life, extending the reach of the \Qbeta{} system well beyond virology.
 
\subsection{Expanding the experimental landscape of viral adaptation}

Recent work has expanded \Qbeta{} evolution studies into extreme environments and biotechnological applications. Studies performed under ionizing radiation, desiccation, simulated microgravity, and in the presence of Martian regolith simulants have opened new opportunities to investigate viral adaptation under physicochemical stresses relevant to astrobiology and planetary exploration~\cite{rodriguez-moreno:2025,arribas-tiemblo:2026,laguna-castro:2024}. In parallel, because the C-terminal extension of A1 projects toward the exterior of the viral capsid~\cite{rumnieks:2011}, it has been successfully exploited for the development of \Qbeta{}-derived display systems capable of exposing foreign peptides on the virion surface while maintaining infectivity ~\cite{singleton:2018,nchinda:2021}. These platforms exploit the same properties that have made \Qbeta{} a powerful experimental model and illustrate how fundamental knowledge gained from studies of viral adaptation can be harnessed for directed evolution and molecular engineering.

\section{Conclusions}

Over more than six decades, bacteriophage \Qbeta{} has repeatedly served as the experimental system in which general principles of molecular evolution were first observed before being recognized more broadly. From Spiegelman's demonstration of Darwinian selection in a cell-free system, through Eigen's formalization of the quasispecies and its first empirical confirmation in \Qbeta{} populations, to the reconstruction of genotype networks by deep sequencing, the phage has consistently offered the resolution and experimental tractability needed to translate theoretical
predictions into direct observation. Complementary studies of adaptation under thermal, ecological, and mutagenic pressures have further established \Qbeta{} as a versatile platform for dissecting how mutation, selection, and population structure jointly shape evolutionary trajectories across molecular, population, and ecological scales.

These contributions extend well beyond the biology of a single bacteriophage. \Qbeta{} replicase has become a building block for synthetic self-replicating systems addressing questions in the origin of life, while the virus itself continues to be pushed into new experimental territories, from extreme, astrobiologically relevant environments to biotechnological display platforms. As high-resolution sequencing and systems-level approaches continue to mature, \Qbeta{} is well positioned to remain a
reference system not only for understanding the evolution of RNA viruses, but for probing the broader principles that govern adaptation in biological populations more generally.

\section{Funding}


This research was supported through grants  PID2023-147963NB-C21 (S.M.), and PID2023-147963NB-C22 (E.L.) funded by MICIU/AEI/10.13039/501100011033 and by European Regional Development Fund, European Union. 

\section{Declaration of Competing Interest}

The authors declare that they have no known competing financial interests or
personal relationships that could have appeared to influence the work reported
in this paper.

\section{Data Availability}

No data were used for the research described in the article.

\section{Declaration of generative AI and AI-assisted technologies in the manuscript preparation process}

During the preparation of this work, the authors used ChatGPT and Claude as support in the bibliographic search and to improve written English. The authors reviewed and edited the output as needed and take full responsibility for the content of the published article. 


%

\IfFileExists{bibliography.bib}{%
  \renewcommand{\bibsection}{%
    \section{References and recommended reading}
    \noindent Papers of particular interest, published within the period of
    review, have been highlighted as:
    \par\smallskip
    \noindent \textbullet\ of special interest
    \par
    \noindent \textbullet\textbullet\ of outstanding interest
  }
  \bibliographystyle{elsarticle-num}
  \bibliography{bibliography}

@STRING{BE = {Behav.\ Ecol.}}

@STRING{PNAS = {Proc.\ Natl.\ Acad.\ Sci.\ USA}}

@STRING{SA = {Sci.\ Adv.}}

@article{acevedo:2014,
 author =  {A Acevedo and L Brodsky and R Andino},
 title =  {Mutational and fitness landscapes of an {RNA} virus
         revealed through population screening},
 journal =  {Nature},
 year =  {2014},
 volume =  {505},
 pages =  {686-690},
}

@article{andino:2015,
title = {Viral quasispecies},
journal = {Virology},
volume = {479-480},
pages = {46-51},
year = {2015},
note = {60th Anniversary Issue},
issn = {0042-6822},
doi = {https://doi.org/10.1016/j.virol.2015.03.022},
url = {https://www.sciencedirect.com/science/article/pii/S0042682215001580},
author = {Raul Andino and Esteban Domingo},
}

@article{arribas:2014,
author = {Arribas, Mar\'{\i}a and Kubota, Kirina and Cabanillas, Laura and L\'azaro, Ester},
year = {2014},
month = {06},
pages = {e100940},
title = {Adaptation to Fluctuating Temperatures in an {RNA} Virus Is Driven by the Most Stringent Selective Pressure},
volume = {9},
journal = {PloS one},
doi = {10.1371/journal.pone.0100940}
}

@article{arribas:2016,
title = {Impact of increased mutagenesis on adaptation to high temperature in bacteriophage {Q$\beta$}},
journal = {Virology},
volume = {497},
pages = {163-170},
year = {2016},
issn = {0042-6822},
doi = {https://doi.org/10.1016/j.virol.2016.07.007},
url = {https://www.sciencedirect.com/science/article/pii/S004268221630174X},
author = {María Arribas and Laura Cabanillas and Kirina Kubota and Ester Lázaro}
}

@article{arribas:2018,
  author = {Arribas, María and Aguirre, Jacobo and Manrubia, Susanna and Lázaro, Ester},
  title = "{Differences in adaptive dynamics determine the success of virus variants that propagate together}",
  journal = {Virus Evolution},
  volume = {4},
  number = {1},
  year = {2018},
  doi = {10.1093/ve/vex043}
}

@article{arribas:2021,
author = {Arribas, Mar\'{\i}a and L\'azaro, Ester},
year = {2021},
month = {06},
pages = {6815},
title = {Intra-Population Competition during Adaptation to Increased Temperature in an {RNA} Bacteriophage},
volume = {22},
journal = {International Journal of Molecular Sciences},
doi = {10.3390/ijms22136815}
}

@article{beerenwinkel:2012,
  author    = {Beerenwinkel, Niko and G{\"u}nthard, Huldrych F. and Roth, Volker and Metzner, Karin J.},
  title     = {Challenges and opportunities in estimating viral genetic diversity from next-generation sequencing data},
  journal   = {Frontiers in Microbiology},
  volume    = {3},
  pages     = {329},
  year      = {2012},
  publisher = {Frontiers Media SA},
  doi       = {10.3389/fmicb.2012.00329}
}

@article{biebricher:1981,
  author    = {Biebricher, Christof K. and Eigen, Manfred and Luce, R\"{u}diger},
  title     = {Kinetic analysis of template-instructed and de novo {RNA} synthesis by {Q$\beta$} replicase},
  journal   = {Journal of Molecular Biology},
  volume    = {148},
  number    = {4},
  pages     = {391--410},
  year      = {1981},
  publisher = {Elsevier}
}

@article{borst:2026,
author = {Piet Borst  and Richard A. Flavell },
title = {Charles {Weissmann} (1931-2025), an outstanding and captivating molecular biologist},
journal = {Proceedings of the National Academy of Sciences},
volume = {123},
number = {15},
pages = {e2606086123},
year = {2026},
doi = {10.1073/pnas.2606086123}
}

@article{domingo:1976,
title = {In vitro site-directed mutagenesis: Generation and properties of an
infectious extracistronic mutant of bacteriophage {Q$\beta$}},
journal = {Gene},
volume = {1},
number = {1},
pages = {3-25},
year = {1976},
issn = {0378-1119},
doi = {https://doi.org/10.1016/0378-1119(76)90003-2},
author = {E. Domingo and R.A. Flavell and C. Weissmann}
}

@article{domingo:1978,
title = {Nucleotide sequence heterogeneity of an {RNA} phage population},
journal = {Cell},
volume = {13},
number = {4},
pages = {735-744},
year = {1978},
issn = {0092-8674},
doi = {https://doi.org/10.1016/0092-8674(78)90223-4},
author = {Esteban Domingo and Donna Sabo and Tadatsugu Taniguchi and Charles Weissmann}
}

@article{domingo:2012,
author = {Esteban Domingo and Julie Sheldon and Celia Perales},
title = {Viral Quasispecies Evolution},
journal = {Microbiology and Molecular Biology Reviews},
volume = {76},
number = {2},
pages = {159-216},
year = {2012}
}

@ARTICLE{eigen:1971,
 author = {Manfred Eigen},
 title = {Selforganization of matter and the evolution of biological macromolecules},
 journal = {Naturwissenschaften},
 year = {1971},
 volume = {58},
 pages = {465--523}
}

@article{flavell:1974,
  author  = {Flavell, R. A. and Sabo, D. L. and Bandle, E. F. and Weissmann, C.},
  title   = {Site-directed mutagenesis: generation of an extracistronic mutation in bacteriophage {Q$\beta$} {RNA}},
  journal = {Journal of Molecular Biology},
  year    = {1974},
  volume  = {89},
  number  = {2},
  pages   = {255--272},
  doi     = {10.1016/0022-2836(74)90517-8}
}

@article{furubayashi:2020,
  author  = {Furubayashi, T. and Ueda, K. and Bansho, Y. and Motooka, D. and Nakamura, S. and Mizuuchi, R. and Ichihashi, N.},
  title   = {Emergence and diversification of a host-parasite {RNA} ecosystem through {D}arwinian evolution},
  journal = {eLife},
  volume  = {9},
  pages   = {e56038},
  year    = {2020},
  doi     = {10.7554/eLife.56038},
  note    = {\specialinterest{This study moves beyond classical \emph{in vitro} evolution experiments to address a central question in origin-of-life research: how complexity can emerge from simple self-replicating molecules. The spontaneous appearance of parasitic {RNA}s and their subsequent coevolution with replicase-encoding host {RNA}s generated sustained diversification and evolutionary arms races, providing experimental support for the idea that host-parasite interactions were important drivers of prebiotic evolution.}}
}

@ARTICLE{grande-perez:2005,
 author = {Ana Grande-P\'erez and Ester L\'azaro and Esteban Domingo and Susanna C. Manrubia},
 title = {Suppression of viral infectivity through lethal defection},
 journal = PNAS,
 year = {2005},
 volume = {102},
 pages = {4448--4452}
}

@article{haruna:1964,
  author    = {Haruna, Ichiro and Spiegelman, Sol},
  title     = {Specific activation of {RNA} replicase by a template {RNA}},
  journal   = {Proceedings of the National Academy of Sciences},
  volume    = {51},
  number    = {1},
  pages     = {56--63},
  year      = {1964},
  publisher = {National Acad Sciences}
}

@Article{hossain:2020,
AUTHOR = {Hossain, Md. Tanvir and Yokono, Toma and Kashiwagi, Akiko},
TITLE = {The Single-Stranded {RNA} Bacteriophage {Q$\beta$} Adapts Rapidly to High Temperatures: An Evolution Experiment},
JOURNAL = {Viruses},
VOLUME = {12},
YEAR = {2020},
NUMBER = {6},
ARTICLE-NUMBER = {638},
URL = {https://www.mdpi.com/1999-4915/12/6/638},
PubMedID = {32545482},
ISSN = {1999-4915},
DOI = {10.3390/v12060638}
}

@article{ichihashi:2013,
  author  = {Ichihashi, N. and Usui, K. and Kazuta, Y. and Sunami, T. and Matsuura, T. and Yomo, T.},
  title   = {Darwinian evolution in a translation-coupled {RNA} replication system within a cell-like compartment},
  journal = {Nature Communications},
  volume  = {4},
  pages   = {2494},
  year    = {2013},
  doi     = {10.1038/ncomms3494}
}

@article{isakov:2015,
  author    = {Isakov, Oded and Border{\'\i}a, Alberto V. and Golan, David and Hamenahem, Avital and Celniker, Gershon and Yoffe, Liat and Blanc, H{\'e}l{\`e}ne and Vignuzzi, Marco and Shomron, Noam},
  title     = {Deep sequencing analysis of viral infection and evolution allows rapid and detailed characterization of viral mutant spectrum},
  journal   = {Bioinformatics},
  volume    = {31},
  number    = {13},
  pages     = {2141--2150},
  year      = {2015},
  publisher = {Oxford University Press},
  doi       = {10.1093/bioinformatics/btv101},
  pmid      = {25701575},
  pmcid     = {PMC4481840}
}

@article{kashiwagi:2014,
author = {Akiko Kashiwagi and Ryu Sugawara and Fumie Sano Tsushima and Tomofumi Kumagai and Tetsuya Yomo and A. Simon },
title = {Contribution of Silent Mutations to Thermal Adaptation of {RNA} Bacteriophage {Q}$\beta$},
journal = {Journal of Virology},
volume = {88},
number = {19},
pages = {11459-11468},
year = {2014},
doi = {10.1128/JVI.01127-14}
}

@Article{kashiwagi:2018,
author={Kashiwagi, Akiko
and Kadoya, Tamami
and Kumasaka, Naoya
and Kumagai, Tomofumi
and Tsushima, Fumie Sano
and Yomo, Tetsuya},
title={Influence of adaptive mutations, from thermal adaptation experiments, on the infection cycle of {RNA} bacteriophage {Q$\beta$}},
journal={Archives of Virology},
year={2018},
month={Oct},
day={01},
volume={163},
number={10},
pages={2655-2662},
issn={1432-8798},
doi={10.1007/s00705-018-3895-6},
url={https://doi.org/10.1007/s00705-018-3895-6}
}

@article{kita:2008,
  author  = {Kita, H. and Matsuura, T. and Sunami, T. and Hosoda, K. and Ichihashi, N. and Tsukada, K. and Urabe, I. and Yomo, T.},
  title   = {Replication of genetic information with self-encoded replicase in liposomes},
  journal = {ChemBioChem},
  volume  = {9},
  number  = {15},
  pages   = {2403--2410},
  year    = {2008},
  doi     = {10.1002/cbic.200800360}
}

@ARTICLE{laguna-castro:2022,
AUTHOR={Laguna-Castro, Mara and L\'azaro, Ester },
TITLE={Propagation of an {RNA} Bacteriophage at Low Host Density Leads to a More Efficient Virus Entry},
JOURNAL={Frontiers in Virology},
VOLUME={2},
YEAR={2022},
URL={https://www.frontiersin.org/journals/virology/articles/10.3389/fviro.2022.858227},
DOI={10.3389/fviro.2022.858227},
SSN={2673-818X}
}

@article{laguna-castro:2023,
AUTHOR={Laguna-Castro, Mara  and Rodríguez-Moreno, Alicia  and Llorente, Elena and Lázaro, Ester },
TITLE={The balance between fitness advantages and costs drives adaptation of bacteriophage {Q$\beta$} to changes in host density at different temperatures},
JOURNAL={Frontiers in Microbiology},
VOLUME={14},
YEAR={2023},
URL={https://www.frontiersin.org/journals/microbiology/articles/10.3389/fmicb.2\
023.1197085},
DOI={10.3389/fmicb.2023.1197085},
ISSN={1664-302X},
NOTE={\specialinterest{This work demonstrates that the same selective challenge (low host density) can be solved through different genetic pathways depending on the environment. The study links host availability, temperature, and fitness trade-offs, illustrating how ecological context shapes adaptive trajectories in {RNA} viruses.}}
}

@Article{laguna-castro:2025,
AUTHOR = {Laguna-Castro, Mara and Somovilla, Pilar and López-Muñoz, Víctor and Pacios, Luis F. and Lázaro, Ester},
TITLE = {Navigating the Fitness Landscape: Host Density, Epistasis, and Clonal Interference Drive Divergent Evolutionary Pathways in Phage {Q$\beta$}},
JOURNAL = {International Journal of Molecular Sciences},
VOLUME = {26},
YEAR = {2025},
NUMBER = {18},
pages = {9020},
URL = {https://www.mdpi.com/1422-0067/26/18/9020},
ISSN = {1422-0067},
DOI = {10.3390/ijms26189020},
}

@article{lazaro:2018,
author = {Lázaro, Ester and Arribas, María and Cabanillas, Laura and Román, Ismael and Acosta, Esther},
year = {2018},
month = {05},
pages = {8080},
title = {Evolutionary adaptation of an {RNA} bacteriophage to the simultaneous increase in the within-host and extracellular temperatures},
volume = {8},
journal = {Scientific Reports},
doi = {10.1038/s41598-018-26443-z}
}

@article{leeks:2023,
  author = {Leeks, Asher and Bono, Lisa M. and Ampolini, Elizabeth A. and Souza, Lucas S. and H\"ofler, Thomas and Mattson, Courtney L. and Dye, Anna E. and D\'iaz-Mu\~noz, Samuel L.},
  title = "{Open questions in the social lives of viruses}",
  journal = {Journal of Evolutionary Biology},
  volume = {36},
  number = {11},
  pages = {1551-1567},
  year = {2023},
  month = {11},
  issn = {1010-061X},
  doi = {10.1111/jeb.14203},
  url = {https://doi.org/10.1111/jeb.14203},
  eprint = {https://academic.oup.com/jeb/article-pdf/36/11/1551/56276186/jeb14203.pdf},
}

@ARTICLE{manrubia:2010,
 author = {Susanna C. Manrubia and Esteban Domingo and Ester L\'azaro},
 title = {Pathways to extinction -- Beyond the error threshold},
 journal = {Phil. Trans. R.. Soc.},
 year = {2010},
 volume = {365},
 pages = {1943--1952}
}

@incollection{manrubia:2026,
author = {Susanna Manrubia and Luis F. Seoane and José A. Cuesta},
title = {The challenge of scale in molecular adaptaton: Local searches in astronomical genotype networks},
booktitle = {Integrative Theory of Evolution -- Aspects and Insights},
publisher = {Springer},
year      = {2026},
editor    = {Guenther Witzany},
pages     = {},
address   = {},
volume    = {},
series    = {},
edition   = {},
doi       = {10.1007/978-3-032-24295-2_26}
}

@article{martinezalcala:2026,
  author  = {Mart{\'i}nez-Alcal{\'a}, Samuel and Atienza-Diez, Iker and
             Somovilla, Pilar and Mart{\'i}nez-Gonz{\'a}lez, Brenda and
             Perales, Celia and Seoane, Luis F. and L{\'a}zaro, Ester and
             Manrubia, Susanna},
  title   = {Shared Quasispecies Architecture in Experimental and Natural {RNA} Virus Populations},
  journal = {Virus Evolution},
  volume = {12},
  issue = {1},
  pages = {veag053},
  year    = {2026},
  doi = {10.1093/ve/veag053},
  eprint  = {2605.13535},
  archivePrefix = {arXiv},
  note = {\specialinterest{This study compares genotype networks reconstructed from {Q$\beta$} populations evolved in the laboratory with those of {SARS-CoV-2} evolving within human hosts. Despite the very different biology, hosts, and selective environments of the two viruses, their networks share key topological features. The results suggest that the network architecture first resolved in {Q$\beta$} may be a generic property of {RNA} virus quasispecies rather than a peculiarity of one experimental system.}}
}

@article{martinezgonzalez:2025,
author = {Brenda Martínez-González  and María Eugenia Soria  and Ana Isabel de Ávila  and Pilar Somovilla  and Claudia Aguilar-Sabido  and Pablo Mínguez  and Cristina Ferrer-Orta  and Llanos Salar-Vidal  and Ramón Lorenzo-Redondo  and Soledad Delgado  and Federico Morán  and Nuria Verdaguer  and Ignacio Gadea  and Esteban Domingo  and Celia Perales },
title = {{SARS-CoV-2} mutant spectrum complexity is an epidemiologically evolvable trait},
journal = {Proceedings of the National Academy of Sciences},
volume = {122},
number = {39},
pages = {e2515706122},
year = {2025},
doi = {10.1073/pnas.2515706122},
URL = {https://www.pnas.org/doi/abs/10.1073/pnas.2515706122},
eprint = {https://www.pnas.org/doi/pdf/10.1073/pnas.2515706122},
note = {\outstandinginterest{Using deep sequencing of {SARS-CoV-2} from infected patients, this work shows that the complexity of the intra-host mutant spectrum is not a mere by-product of replication. It varies systematically with host and pathogen factors and behaves as a trait that can evolve at the epidemiological scale. The study brings the quasispecies concept, long considered mostly in experimental systems, into the analysis of natural infections of medical relevance.}}
}

@article{mills:1967,
  author  = {Mills, D. R. and Peterson, R. L. and Spiegelman, S.},
  title   = {An Extracellular {D}arwinian Experiment with a Self-Duplicating Nucleic Acid Molecule},
  journal = {Proceedings of the National Academy of Sciences of the USA},
  year    = {1967},
  volume  = {58},
  number  = {1},
  pages   = {217--224},
  doi     = {10.1073/pnas.58.1.217}
}

@article{mizuuchi:2018,
  author  = {Mizuuchi, R. and Ichihashi, N.},
  title   = {Sustainable replication and coevolution of cooperative {RNA}s in an artificial cell-like system},
  journal = {Nature Ecology \& Evolution},
  volume  = {2},
  number  = {10},
  pages   = {1654--1660},
  year    = {2018},
  doi     = {10.1038/s41559-018-0650-z}
}

@article{oehlenschlager:1997,
  author  = {Oehlenschl{\"a}ger, Frank and Eigen, Manfred},
  title   = {30 Years Later -- A New Approach to Sol Spiegelman's and Leslie Orgel's in vitro Evolutionary Studies. Dedicated to Leslie Orgel on the Occasion of His 70th Birthday},
  journal = {Origins of Life and Evolution of the Biosphere},
  year    = {1997},
  month   = dec,
  volume  = {27},
  number  = {5-6},
  pages   = {437--457},
  doi     = {10.1023/A:1006501326129},
  pmid    = {9394469}
}

@article{saffhill:1970,
  author = {Saffhill, R. and Schneider-Bernloehr, H. and Orgel, Leslie E. and Spiegelman, Sol},
  title = {In Vitro Selection of Bacteriophage {Q$\beta$} Ribonucleic Acid Variants Resistant to Ethidium Bromide},
  journal = {Journal of Molecular Biology},
  year = {1970},
  volume = {51},
  number = {3},
  pages = {531--539},
  doi = {10.1016/0022-2836(70)90006-9}
}

@article{salmond:2015,
  title={A century of the phage: past, present and future},
  author={Salmond, George PC and Fineran, Philip C},
  journal={Nature Reviews Microbiology},
  volume={13},
  number={12},
  pages={777--786},
  year={2015},
  publisher={Nature Publishing Group},
  doi={10.1038/nrmicro3564}
}

@article{sakurai:1968,
  author  = {Sakurai, T. and Miyake, T. and Shiba, T. and Watanabe, I.},
  title   = {Isolation of a possible fourth group of {RNA} phage},
  journal = {Japanese Journal of Microbiology},
  volume  = {12},
  number  = {4},
  pages   = {544--546},
  year    = {1968},
  doi     = {10.1111/j.1348-0421.1968.tb00429.x}
}

@article{schuster:1999,
  author    = {Schuster, Peter and Fontana, Walter},
  title     = {Chance and necessity in evolution: lessons from {RNA}},
  journal   = {Physica D: Nonlinear Phenomena},
  volume    = {133},
  number    = {1-4},
  pages     = {427--452},
  year      = {1999},
  publisher = {Elsevier},
  doi       = {10.1016/S01672789(99)00076-6}
}

@article{seoane:2026,
author = {Luís F. Seoane  and Henry Secaira-Morocho  and Pilar Somovilla  and Ester Lázaro  and Susanna Manrubia },
title = {Hierarchical genotype networks and incipient ecological speciation in {Q$\beta$} phage quasispecies},
journal = {Proceedings of the National Academy of Sciences},
volume = {123},
number = {14},
pages = {e2512150123},
year = {2026},
doi = {10.1073/pnas.2512150123},
URL = {https://www.pnas.org/doi/abs/10.1073/pnas.2512150123},
eprint = {https://www.pnas.org/doi/pdf/10.1073/pnas.2512150123},
note = {\outstandinginterest{This study uses deep sequencing of the A2-coding region to reconstruct genotype networks of {Q$\beta$} populations, each comprising tens of thousands of haplotypes linked by single mutations. Across populations and environments, the networks show a hierarchical organization: the population exhaustively samples the mutational neighborhood of its dominant sequence while keeping a periphery of rare variants. This turns the classical mutation--selection balance into a dynamic, structured process around a shifting master sequence.}}
}

@Article{somovilla:2019,
AUTHOR = {Somovilla, Pilar and Manrubia, Susanna and L\'azaro, Ester},
TITLE = {Evolutionary Dynamics in the {RNA} Bacteriophage {Q}$\beta$ Depends on the Pattern of Change in Selective Pressures},
JOURNAL = {Pathogens},
VOLUME = {8},
YEAR = {2019},
NUMBER = {2},
ARTICLE-NUMBER = {80},
DOI = {10.3390/pathogens8020080}
}

@Article{somovilla:2022,
AUTHOR = {Somovilla, Pilar and Rodr\'{\i}guez-Moreno, Alicia and Arribas, Mar\'{\i}a and Manrubia, Susanna and L\'azaro, Ester},
TITLE = {Standing Genetic Diversity and Transmission Bottleneck Size Drive Adaptation in Bacteriophage {Q}$\beta$},
JOURNAL = {International Journal of Molecular Sciences},
VOLUME = {23},
YEAR = {2022},
NUMBER = {16},
ARTICLE-NUMBER = {8876},
DOI = {10.3390/ijms23168876}
}

@article{spiegelman:1965,
  author  = {Spiegelman, S. and Haruna, I. and Holland, I. B. and Beaudreau, G. and Mills, D.},
  title   = {The Synthesis of a Self-Propagating and Infectious Nucleic Acid with a Purified Enzyme},
  journal = {Proceedings of the National Academy of Sciences of the USA},
  year    = {1965},
  volume  = {54},
  number  = {4},
  pages   = {919--927},
  doi     = {10.1073/pnas.54.3.919}
}

@article{swanstrom:2022,
author = {Ronald Swanstrom  and Raymond F. Schinazi },
title = {Lethal mutagenesis as an antiviral strategy},
journal = {Science},
volume = {375},
number = {6580},
pages = {497-498},
year = {2022},
doi = {10.1126/science.abn0048},
URL = {https://www.science.org/doi/abs/10.1126/science.abn0048}
}

@Article{vignuzzi:2006,
author={Vignuzzi, Marco
and Stone, Jeffrey K.
and Arnold, Jamie J.
and Cameron, Craig E.
and Andino, Raul},
title={Quasispecies diversity determines pathogenesis through cooperative interactions in a viral population},
journal={Nature},
year={2006},
month={Jan},
day={01},
volume={439},
number={7074},
pages={344-348},
issn={1476-4687},
doi={10.1038/nature04388},
url={https://doi.org/10.1038/nature04388}
}

@article{weissmann:2011,
  author    = {Weissmann, Charles},
  title     = {The end of the road},
  journal   = {Prion},
  volume    = {5},
  number    = {4},
  pages     = {233--243},
  year      = {2011},
  publisher = {Taylor \& Francis}
}

@ARTICLE{wilke:2005,
 author = {C. O. Wilke},
 title = {Quasispecies theory in the context of population genetics},
 journal = {BMC Evol. Biol.},
 year = {2005},
 volume = {5},
 pages = {44}
}

@article{arribas-tiemblo:2026,
  author = {Arribas Tiemblo, Miguel and Rodr{\'i}guez-Moreno, Alicia and G{\'o}mez, Felipe and L{\'a}zaro, Ester},
  title = {Regolith as a Refuge: Differential Survival of Bacteriophage {Q$\beta$} in Mars Analog Environments},
  journal = {Astrobiology},
  year = {2026},
  volume = {26},
  number = {6},
  pages = {443--457},
  doi = {10.1177/15311074261459203}
}

@article{bernhardt:2001,
  author = {Bernhardt, Thomas G. and Wang, I.-N. and Struck, Douglas K. and Young, Ry},
  title = {A Protein Antibiotic in the Phage {Q$\beta$} Virion: Diversity in Lysis Targets},
  journal = {Science},
  year = {2001},
  volume = {292},
  number = {5525},
  pages = {2326--2329},
  doi = {10.1126/science.1058289}
}

@incollection{biebricher:2006,
  author = {Biebricher, Christof K. and Eigen, Manfred},
  title = {What Is a Quasispecies?},
  booktitle = {Quasispecies: Concept and Implications for Virology},
  series = {Current Topics in Microbiology and Immunology},
  year = {2006},
  volume = {299},
  pages = {1--31},
  publisher = {Springer},
  doi = {10.1007/3-540-26397-7_1}
}

@article{blumenthal:1979,
  author = {Blumenthal, Thomas and Carmichael, Gordon G.},
  title = {{RNA} Replication: Function and Structure of {Q$\beta$} Replicase},
  journal = {Annual Review of Biochemistry},
  year = {1979},
  volume = {48},
  pages = {525--548},
  doi = {10.1146/annurev.bi.48.070179.002521}
}

@article{bradwell:2013,
  author = {Bradwell, Katie and Combe, Marine and Domingo-Calap, Pilar and Sanju{\'a}n, Rafael},
  title = {Correlation between Mutation Rate and Genome Size in Riboviruses: Mutation Rate of Bacteriophage {Q$\beta$}},
  journal = {Genetics},
  year = {2013},
  volume = {195},
  number = {1},
  pages = {243--251},
  doi = {10.1534/genetics.113.154963}
}

@article{bull:2004,
  author = {Bull, James J. and Badgett, Michael R. and Springman, Rachel and Molineux, Ian J.},
  title = {Genome Properties and the Limits of Adaptation in Bacteriophages},
  journal = {Evolution},
  year = {2004},
  volume = {58},
  number = {4},
  pages = {692--701},
  doi = {10.1111/j.0014-3820.2004.tb00402.x}
}

@article{cabanillas:2013,
  author = {Cabanillas, Laura and Arribas, Mar{\'i}a and L{\'a}zaro, Ester},
  title = {Evolution at Increased Error Rate Leads to the Coexistence of Multiple Adaptive Pathways in an {RNA} Virus},
  journal = {BMC Evolutionary Biology},
  year = {2013},
  volume = {13},
  pages = {11},
  doi = {10.1186/1471-2148-13-11}
}

@article{cabanillas:2014,
  author = {Cabanillas, Laura and Sanju{\'a}n, Rafael and L{\'a}zaro, Ester},
  title = {Changes in Protein Domains outside the Catalytic Site of the Bacteriophage {Q$\beta$} Replicase Reduce the Mutagenic Effect of 5-Azacytidine},
  journal = {Journal of Virology},
  year = {2014},
  volume = {88},
  number = {18},
  pages = {10480--10487},
  doi = {10.1128/JVI.00979-14}
}

@article{cases-gonzalez:2008,
  author = {Cases-Gonz{\'a}lez, Carlos and Arribas, Mar{\'i}a and Domingo, Esteban and L{\'a}zaro, Ester},
  title = {Beneficial Effects of Population Bottlenecks in an {RNA} Virus Evolving at Increased Error Rate},
  journal = {Journal of Molecular Biology},
  year = {2008},
  volume = {384},
  number = {5},
  pages = {1120--1129},
  doi = {10.1016/j.jmb.2008.10.014}
}

@article{cui:2017,
  author = {Cui, Zheng and Gorzelnik, Kristin V. and Chang, Ji Young and Langlais, Christian and Jakana, Joanita and Young, Ry and Zhang, Junjie},
  title = {Structures of {Q$\beta$} Virions, Virus-Like Particles, and the {Q$\beta$--MurA} Complex Reveal Internal Coat Proteins and the Mechanism of Host Lysis},
  journal = {Proceedings of the National Academy of Sciences of the USA},
  year = {2017},
  volume = {114},
  number = {44},
  pages = {11697--11702},
  doi = {10.1073/pnas.1707102114}
}

@article{domingo-calap:2010,
  author = {Domingo-Calap, Pilar and Pereira-G{\'o}mez, Marcos and Sanju{\'a}n, Rafael},
  title = {Selection for Thermostability Can Lead to the Emergence of Mutational Robustness in an {RNA} Virus},
  journal = {Journal of Evolutionary Biology},
  year = {2010},
  volume = {23},
  number = {11},
  pages = {2453--2460},
  doi = {10.1111/j.1420-9101.2010.02107.x}
}

@incollection{domingo:2016,
  author = {Domingo, Esteban and Schuster, Peter},
  title = {What Is a Quasispecies? {Historical} Origins and Current Scope},
  booktitle = {Viral Quasispecies: Current Topics in Microbiology and Immunology},
  year = {2016},
  volume = {392},
  pages = {1--22},
  publisher = {Springer},
  doi = {10.1007/82_2015_453}
}

@article{eigen-schuster:1977,
  author = {Eigen, Manfred and Schuster, Peter},
  title = {The Hypercycle: A Principle of Natural Self-Organization. Part A: Emergence of the Hypercycle},
  journal = {Naturwissenschaften},
  year = {1977},
  volume = {64},
  number = {11},
  pages = {541--565},
  doi = {10.1007/BF00450633}
}

@article{garcia-villada:2013,
  author = {Garc{\'i}a-Villada, Libertad and Drake, John W.},
  title = {Experimental Selection Reveals a Trade-Off between Fecundity and Lifespan in the Coliphage {Q$\beta$}},
  journal = {Open Biology},
  year = {2013},
  volume = {3},
  number = {6},
  pages = {130043},
  doi = {10.1098/rsob.130043}
}

@article{haruna:1965,
  author = {Haruna, Ichiro and Spiegelman, Sol},
  title = {Autocatalytic Synthesis of a Viral {RNA} in Vitro},
  journal = {Science},
  year = {1965},
  volume = {150},
  number = {3698},
  pages = {884--886},
  doi = {10.1126/science.150.3698.884}
}

@article{hofstetter:1974,
  author = {Hofstetter, Heinz and Monstein, Hans-J{\"u}rg and Weissmann, Charles},
  title = {The Readthrough Protein {A1} Is Essential for the Formation of Viable {Q$\beta$} Particles},
  journal = {Biochimica et Biophysica Acta},
  year = {1974},
  volume = {374},
  number = {2},
  pages = {238--251},
  doi = {10.1016/0005-2787(74)90366-9}
}

@article{kashiwagi:2011,
  author = {Kashiwagi, Akiko and Yomo, Tetsuya},
  title = {Ongoing Phenotypic and Genomic Changes in Experimental Coevolution of {RNA} Bacteriophage {Q$\beta$} and {Escherichia coli}},
  journal = {PLoS Genetics},
  year = {2011},
  volume = {7},
  number = {8},
  pages = {e1002188},
  doi = {10.1371/journal.pgen.1002188}
}

@article{laguna-castro:2024,
  author = {Laguna-Castro, Mara and Rodr{\'i}guez-Moreno, Alicia and L{\'a}zaro, Ester},
  title = {Evolutionary Adaptation of an {RNA} Bacteriophage to Repeated Freezing and Thawing Cycles},
  journal = {International Journal of Molecular Sciences},
  year = {2024},
  volume = {25},
  number = {9},
  pages = {4863},
  doi = {10.3390/ijms25094863}
}

@article{nchinda:2021,
  author = {Nchinda, Godwin W. and Al-Atoom, Naser and Coats, Maria T. and Cameron, James M. and Waffo, Alain B.},
  title = {Uniqueness of {RNA} Coliphage {Q$\beta$} Display System in Directed Evolutionary Biotechnology},
  journal = {Viruses},
  year = {2021},
  volume = {13},
  number = {4},
  pages = {568},
  doi = {10.3390/v13040568},
  note = {\specialinterest{This review illustrates the transition of {Q$\beta$} from an experimental model of viral evolution to a platform for directed evolutionary biotechnology. The text describes how the phage's quasispecies nature, error-prone replication, and A1-based display system provide advantages for peptide evolution and molecular engineering applications.}}
}

@article{rodriguez-moreno:2025,
  author = {Rodr{\'i}guez-Moreno, Alicia and Mart{\'i}n-Bl{\'a}zquez, Sergio and L{\'o}pez de Heredia, Unai and Soto, {\'A}lvaro and L{\'a}zaro, Ester},
  title = {Impact of Simulated Microgravity in Short-Term Evolution of an {RNA} Bacteriophage},
  journal = {Frontiers in Microbiology},
  year = {2025},
  volume = {16},
  pages = {1680651},
  doi = {10.3389/fmicb.2025.1680651}
}

@article{rumnieks:2011,
  author = {Rumnieks, Janis and Tars, Kaspars},
  title = {Crystal Structure of the Read-Through Domain from Bacteriophage {Q$\beta$} {A1} Protein},
  journal = {Protein Science},
  year = {2011},
  volume = {20},
  number = {10},
  pages = {1707--1712},
  doi = {10.1002/pro.704}
}

@article{singleton:2018,
  author = {Singleton, Ryan L. and Sanders, Crystal A. and Jones, Kimberly and Thorington, Benjamin and Egbo, Temitope and Coats, Maria T. and Waffo, Alain B.},
  title = {Function of the {RNA} Coliphage {Q$\beta$} Proteins in Medical In Vitro Evolution},
  journal = {Methods and Protocols},
  year = {2018},
  volume = {1},
  number = {2},
  pages = {18},
  doi = {10.3390/mps1020018}
}

@article{vignuzzi-lopez:2019,
  author = {Vignuzzi, Marco and L{\'o}pez, Carolina B.},
  title = {Defective Viral Genomes Are Key Drivers of the Virus--Host Interaction},
  journal = {Nature Microbiology},
  year = {2019},
  volume = {4},
  number = {7},
  pages = {1075--1087},
  doi = {10.1038/s41564-019-0465-y}
}

@article{walker:2021,
  author = {Walker, Peter J. and Siddell, Stuart G. and Lefkowitz, Elliot J. and
            Mushegian, Arcady R. and Adriaenssens, Evelien M. and
            Alfenas-Zerbini, Poliane and Davison, Andrew J. and
            Dempsey, Donald M. and Dutilh, Bas E. and
            Garc{\'i}a, Mar{\'i}a Laura and Harrach, Bal{\'a}zs and
            Harrison, Robert L. and Hendrickson, R. Curtis and
            Junglen, Sandra and Knowles, Nick J. and Krupovic, Mart and
            Kuhn, Jens H. and Lambert, Amy J. and
            {\L}obocka, Ma{\l}gorzata and Nibert, Max L. and
            Oksanen, Hanna M. and Orton, Richard J. and
            Robertson, David L. and Rubino, Luisa and
            Sabanadzovic, Sead and Simmonds, Peter and Smith, Donald B. and
            Suzuki, Nobuhiro and Van Dooerslaer, Koenraad and
            Vandamme, Anne-Mieke and Varsani, Arvind and
            Zerbini, Francisco Murilo},
  title = {Changes to Virus Taxonomy and to the International Code of Virus Classification and Nomenclature Ratified by the International Committee on Taxonomy of Viruses (2021)},
  journal = {Archives of Virology},
  year = {2021},
  volume = {166},
  number = {9},
  pages = {2633--2648},
  doi = {10.1007/s00705-021-05156-1}
}

@article{watanabe:1967,
  author = {Watanabe, Ichiro and Miyake, T. and Sakurai, T. and Shiba, T. and Ohno, T.},
  title = {Isolation and Grouping of {RNA} Phages},
  journal = {Proceedings of the Japan Academy},
  year = {1967},
  volume = {43},
  pages = {204--209}
}
}{%
  \section{References and recommended reading}
  \noindent Papers of particular interest, published within the period of
  review, should be highlighted as being of special or outstanding interest.

  [Add your existing bibliography.bib file to this directory.]
}

\end{document}